\documentclass[12]{article}
\usepackage{fleqn}
\usepackage{graphicx}
\usepackage{amssymb}
\begin{document}
\Large
\begin{center}
{\bf On the Veldkamp Space of GQ(4,\,2)}
\end{center}
\large
\vspace*{.0cm}
\begin{center}
Metod Saniga
\end{center}
\vspace*{-.5cm} \normalsize
\begin{center}
Astronomical Institute, Slovak Academy of Sciences\\
SK-05960 Tatransk\' a Lomnica, Slovak Republic\\
(msaniga@astro.sk)

\vspace*{.2cm}

(16 February 2010)

\end{center}

\vspace*{-.3cm} \noindent \hrulefill

\vspace*{.0cm} \noindent {\bf Abstract}

\noindent The Veldkamp space, in the sense of Buekenhout and
Cohen, of the generalized quadrangle GQ(4,\,2) is shown not to be
a (partial) linear space by simply giving several examples of
Veldkamp lines (V-lines) having two or even three Veldkamp points
(V-points) in common. Alongside the ordinary V-lines of size five,
one also finds V-lines of cardinality three and two. There,
however, exists a subspace of the Veldkamp space isomorphic to
PG(3,\,4) having 45 perps and 40 plane ovoids as its 85 V-points,
with its 357 V-lines being of four distinct types. A V-line of the
first type consists of five perps on a common line (altogether 27
of them),  the second type features three perps and two ovoids
sharing a tricentric triad (240 members), whilst the third and
fourth type each comprises a perp and four ovoids in the rosette
centered at the (common) center of the perp (90).  It is also
pointed out that 160 non-plane ovoids (tripods) fall into two
distinct orbits --- of sizes 40 and 120 --- with respect to the
stabilizer group of a copy of GQ(2,\,2); a tripod of the
first/second orbit sharing with the GQ(2,\,2) a
tricentric/unicentric triad, respectively.
Finally, three remarkable subconfigurations of V-lines represented
by fans of ovoids through a fixed ovoid are examined in some detail.
\\ \\
%{\bf MSC Codes:} 51Exx, 81R99~~~~~~~~~~~~~~~~~
%{\bf PACS Numbers:} 02.10.Ox, 02.40.Dr, 03.65.Ca\\

\vspace*{-.2cm} \noindent \hrulefill

\vspace*{.3cm}
%\large
\section{Introduction}
Generalized quadrangles of types GQ($2, t$), with $t$ = 1, 2, and
4, have recently been found to play a prominent role in quantum
information and black hole physics; the first type for grasping
the geometrical nature of the so-called Mermin squares
\cite{qic,tmp}, the second for underlying commutation properties
between the elements of two-qubit Pauli group \cite{qic,tmp,hos},
and the third one for fully encoding the $E_{6(6)}$ symmetric
entropy formula describing black holes and black strings in $D=5$
\cite{gq24}. Whereas GQ(2,\,2) is isomorphic to its point-line
dual, this is not the case with the remaining two geometries; the
dual of GQ(2,\,1) being GQ(1,\,2), that of GQ(2,\,4) GQ(4,\,2)
\cite{paythas}. These two duals, strangely, did not appear in the
above-mentioned physical contexts.  It is, therefore, natural to
ask why this is so. We shall try to shed light on this matter by
invoking the concept of the Veldkamp space of a point line
incidence structure \cite{buek}.  The Veldkamp space of GQ(2,\,4)
was shown to be a linear space isomorphic to PG(5,\,2)
\cite{vs24}.  Here, we shall demonstrate that the Veldkamp space
of GQ(4,\,2), due to the existence of two different kinds of
ovoids in GQ(4,\,2), is not even a partial linear space, though it
contains a (linear) subspace isomorphic to PG(3,\,4) in which
GQ(4,\,2) lives as a non-degenerate  Hermitian variety.

\section{Rudiments of the Theory of Finite Generalized Quadrangles and
    Veldkamp Spaces}

To make our exposition as self-contained as possible, and for the
reader's convenience as well, we will first gather all essential
information about finite generalized quadrangles \cite{paythas},
then introduce the concept of a geometric hyperplane \cite{ron}
and, finally, that of the Veldkamp space of a point-line incidence
geometry \cite{buek}.

A {\it finite generalized quadrangle} of order $(s, t)$, usually
denoted GQ($s, t$), is an incidence structure $S = (P, B, {\rm
I})$, where $P$ and $B$ are disjoint (non-empty) sets of objects,
called respectively points and lines, and where I is a symmetric
point-line incidence relation satisfying the following axioms
\cite{paythas}: (i) each point is incident with $1 + t$ lines ($t
\geq 1$) and two distinct points are incident with at most one
line; (ii) each line is incident with $1 + s$ points ($s \geq 1$)
and two distinct lines are incident with at most one point;  and
(iii) if $x$ is a point and $L$ is a line not incident with $x$,
then there exists a unique pair $(y, M) \in  P \times B$ for which
$x {\rm I} M {\rm I} y {\rm I} L$; from these axioms it readily
follows that $|P| = (s+1)(st+1)$ and $|B| = (t+1)(st+1)$. It is
obvious that there exists a point-line duality with respect to
which each of the axioms is self-dual. Interchanging points and
lines in $S$ thus yields a generalized quadrangle $S^{D}$ of order
$(t, s)$, called the dual of $S$. If $s = t$, $S$ is said to have
order $s$. The generalized quadrangle of order $(s, 1)$ is called
a grid and that of order $(1, t)$ a dual grid. A generalized
quadrangle with both $s > 1$ and $t > 1$ is called thick. Every
finite generalized quadrangle is obviously a {\it partial} linear
space, that is the point-line incidence structure where a) any
line has at least two points and b) any two points are on at most
one line.

Given two points $x$ and $y$ of $S$ one writes $x \sim y$ and says
that $x$ and $y$ are collinear if there exists a line $L$ of $S$
incident with both. For any $x \in P$ define the perp of $x$ as
$x^{\perp} = \{y \in P | y \sim x \}$ and note that $x \in
x^{\perp}$, being its center;  obviously, $|x^{\perp}| = 1+s+st$.
Given an arbitrary subset $A$ of $P$, the {\it perp} of $A$,
$A^{\perp}$, is defined as $A^{\perp} = \bigcap \{x^{\perp} | x
\in A\}$ and $A^{\perp \perp} := (A^{\perp})^{\perp}$; in
particular, if $x$ and $y$ are two non-collinear points, then
$\{x,y \}^{\perp \perp}$ is called a hyperbolic line (through
them). A triple of pairwise non-collinear points of $S$ is called
a {\it triad}; given any triad $T$, a point of $T^{\perp}$ is
called its center and we say that $T$ is acentric, centric or
unicentric according as $|T^{\perp}|$ is, respectively, zero,
non-zero or one. An ovoid of a generalized quadrangle $S$ is a set
of points of $S$ such that each line of $S$ is incident with
exactly one point of the set;  hence, each ovoid contains $st + 1$
points. The dual concept is that of spread; this is a set of lines
such that every point of $S$ is on a unique line of the spread. A
{\it rosette} of ovoids is a set of ovoids through a given point
$x$ of $S$ partitioning the set of points non-collinear with $x$.
A {\it fan} of ovoids is a set of ovoids partitioning the whole
point set of $S$; if $S$ has order $(s,\,t)$ then every rosette
contains $s$ ovoids and every fan features $s + 1$ ovoids.

A {\it geometric hyperplane} $H$ of a point-line geometry $\Gamma
(P,B)$ is a proper subset of $P$ such that each line of $\Gamma$
meets $H$ in one or all points \cite{ron}. For $\Gamma =$ GQ($s,
t$), it is well known that $H$ is one of the following three kinds
\cite{paythas}: (i) the perp of a point $x$,  $x^{\perp}$; (ii) a
(full) subquadrangle of order ($s,t'$), $t' < t$; and (iii) an
ovoid.

Finally, we shall introduce the notion of the {\it Veldkamp space}
of a point-line incidence geometry $\Gamma(P,B)$,
$\mathcal{V}(\Gamma)$ \cite{buek}. $\mathcal{V}(\Gamma)$  is the
space in  which (i) a point is a geometric hyperplane of  $\Gamma$
and (ii) a line is the collection $H_{1}H_{2}$ of all geometric
hyperplanes $H$ of $\Gamma$  such that $H_{1} \cap H_{2} = H_{1}
\cap H = H_{2} \cap H$ or $H = H_{i}$ ($i = 1, 2$), where $H_{1}$
and  $H_{ 2}$ are  distinct points of $\mathcal{V}(\Gamma)$.

\section{Basic Properties of GQ(4,\,2)}
The unique generalized quadrangle GQ(4,\,2), associated with the
classical group PGU$_{4}$(2), can be represented by 45 points and
27 lines of a non-degenerate Hermitian surface $H$(3,\,4) in
PG(3,\,4), the three-dimensional projective space over GF(4)
\cite{paythas,bw,hir}. Every line has five points and there are
three lines through every point. This quadrangle features both
unicentric and tricentric triads,\footnote{GQ(4,\,2) is also
endowed with acentric triads, but these are of no relevance for
our subsequent reasoning.}  and has no spreads. There are 16
tricentric triads through each point; hence, their total number is
$45 \times 16 / 3 = 240.$ Its geometric hyperplanes are (45) perps
of points and (200) ovoids, because it has no subquadrangles of
type GQ(4,\,1) \cite{paythas}.

Obviously, perps correspond to the cuts of $H(3,\,4)$  by its 45
tangent planes. As first shown by Brouwer and Wilbrink \cite{bw},
ovoids fall into two distinct orbits of sizes 40 and 160.  The
ovoids of the first orbit are called {\it plane} ovoids, as each
of them represents a section of $H(3,\,4)$ by one of the 40
non-tangent planes.  The ovoids of the second orbit are referred
to as {\it tripods}, each comprising nine isotropic points on
three hyperbolic lines $\{x, y \}^{\perp \perp}$, $\{x, z
\}^{\perp \perp}$ and $\{x, w \}^{\perp \perp}$, where $\{x, y, z,
w \}$ is a basis of non-isotropic points; in other words, every tripod
can be viewed as a unique union of three tricentric triads. Given a plane ovoid
$\mathcal{P}$ and any two distinct points $x, y \in \mathcal{P}$,
it is {\it always} true that $\{x, y \}^{\perp \perp} \subseteq
\mathcal{P}$. Hence,  $\{ \mathcal{P} \setminus \{x, y \}^{\perp
\perp}  \} \cup \{x, y \}^{\perp} $ is again an ovoid, and all the
tripods can be obtained in this manner from plane ovoids.
GQ(4,\,2) contains both fans and rosettes of ovoids \cite{bw}.
There are altogether 520 fans, each made of a plane ovoid and four
tripods and falling into two orbits of sizes 480 and 40, and 26
rosettes on a given point; two of them feature four plane ovoids,
the remaining ones consist of four tripods each.

Apart from perps and ovoids, GQ(4,\,2) is endowed with one more
kind of distinguished subgeometry --- that isomorphic to the
unique generalized quadrangle GQ(2,\,2); this is, however, not a
geometric hyperplane. There are altogether 36 distinct copies of
GQ(2,\,2) living inside GQ(4,\,2), and with any of them an ovoid
is found to share a triad.  For a plane ovoid this triad is always
tricentric. For tripods, however, it can be either tricentric (40
of them) or unicentric (120 of them); in what follows we shall
occasionally refer to the former/latter as
tri-tripods/uni-tripods, respectively.
%Two plane ovoids (tri-tripods) overlap in either a single point or a
%triad. Given a plane ovoid (tri-tripod), there are 10 tri-tripods
%(plane ovoids) and 30 (21) uni-tripods disjoint from it; given a
%uni-tripod, there are 10 plane ovoids, 7 tri-tripods and 14 uni-tripods
%disjoint from it.
There exists a remarkable partitioning of the point-set of
GQ(4,\,2) in terms of three GQ(2,\,1)s and three GQ(1,2)s such
that one of the latter group forms with each of the former group a
GQ(2,\,2). Another noteworthy property is the existence of pairs
of plane ovoids and/or tri-tripods on the common (tricentric)
triad whose symmetric difference is a disjoint union of two
GQ(1,\,2)s.

\begin{figure}[t]\unitlength1.0cm%
\centering
  \begin{picture}(8.4,7.8)%%  8.4 wide and 7.8 high
    \put(0,0){\includegraphics[height=8\unitlength]{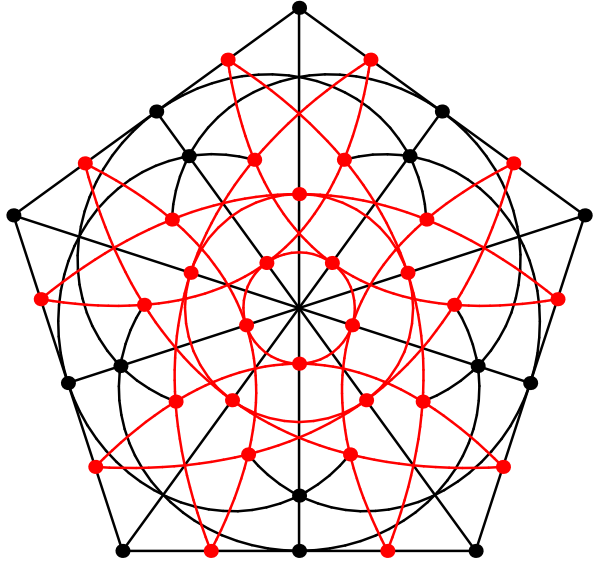}}
    % \put(3.375,0){$A$}
  \end{picture}
    \caption{A diagrammatical model of the structure of GQ(4,\,2) whose points
    are illustrated by bullets and lines by straight segments, arcs of ellipses and/or
    parabolas, and two circles (for more details, see the text). Note a particular
    copy of GQ(2,\,2) (black), its complement (red) being nothing but famous Schl\" afli's double-six of lines.}
\end{figure}

All the above-mentioned properties can be ascertained --- some
readily, some requiring a bit of work --- from a diagrammatical
illustration of GQ(4,\,2) depicted in Figure 1. In this picture,
of a form showing an automorphism of order five, all the 45 points
of GQ(4,\,2) are represented by bullets, whereas its 27 lines have
as many as four distinct representations: two are represented by
(concentric) circles, five by arcs of parabolas touching the inner
circle, another five by parabolas touching the outer circle, five
by arcs of ellipses, and, finally, 10 by straight line-segments.
(Note that there are many intersections of segments and arcs that
do not stand for any point of GQ(4,\,2).) A copy of GQ(2,\,2) is
also highlighted (black bullets, all line-segments and all arcs of
ellipses).

\section{Distinguished Features of the Veldkamp Space of GQ(4,\,2)}
\subsection{Linear Subspace Isomorphic to PG(3,\,4)}
In PG(3,\,4) a point and a plane are duals of each other. On the
other hand, both a perp and a plane ovoid are associated each with
a unique plane of PG(3,\,4). Hence, disregarding tripods for the
moment,  we find a subspace of the Veldkamp space of GQ(4,\,2)
that is isomorphic to PG(3,\,4). The 85 V-points of this subspace
are 45 perps and 40 planar ovoids, and the 357 V-lines split into
four distinct types as shown in Figure 2. A V-line of the first
type (1-st row in Figure 2) consists of five perps on a common
pentad of collinear points i.\,e. on a common line; clearly, there
are 27 V-lines of this type as each line leads to a unique V-line.
A second-type V-line (2-nd row) features three perps and two
ovoids sharing a tricentric triad; since each such triad defines a
unique V-line, there are altogether 240 V-lines of this type.
Third-/fourth-type V-lines (3-rd/4-th row) each comprises a perp
and four ovoids in the rosette centered at the perp's center (the
only common point); their total number thus amounts to  $2 \times
45 =  90$.

\begin{figure}[!h]\unitlength0.35cm%%  8.4 breit und 7.8 hoch
\centering
  %{}\hfill
  \begin{picture}(8.4,7.8)%%  8.4 wide and 7.8 high
    \put(0,0){\includegraphics[height=8\unitlength]{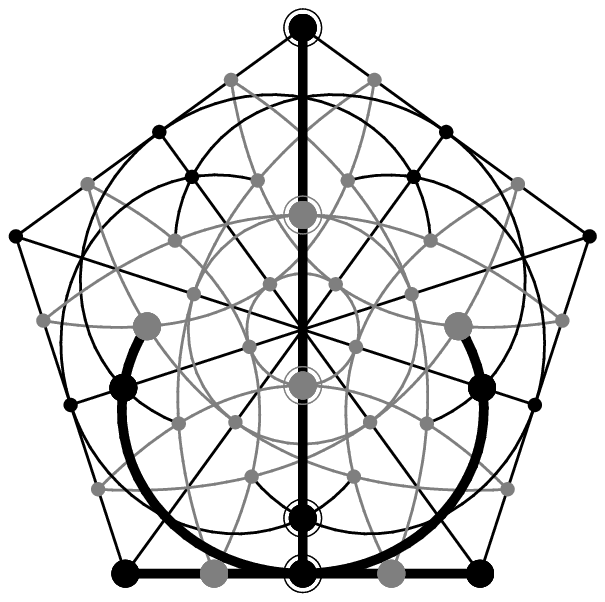}}
    % \put(3.375,0){$A$}
  \end{picture}%
  \hfill
  \begin{picture}(8.4,7.8)%%  8.4 wide and 7.8 high
    \put(0,0){\includegraphics[height=8\unitlength]{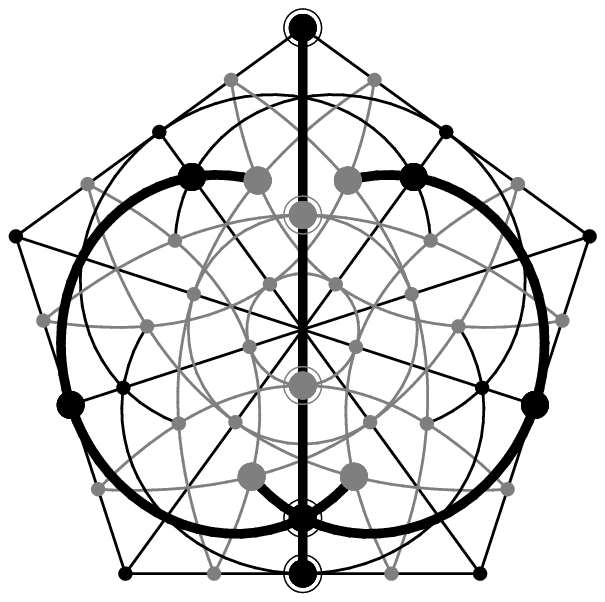}}
    % \put(3.375,0){$A$}
  \end{picture}%
  \hfill
  \begin{picture}(8.4,7.8)%%  8.4 wide and 7.8 high
    \put(0,0){\includegraphics[height=8\unitlength]{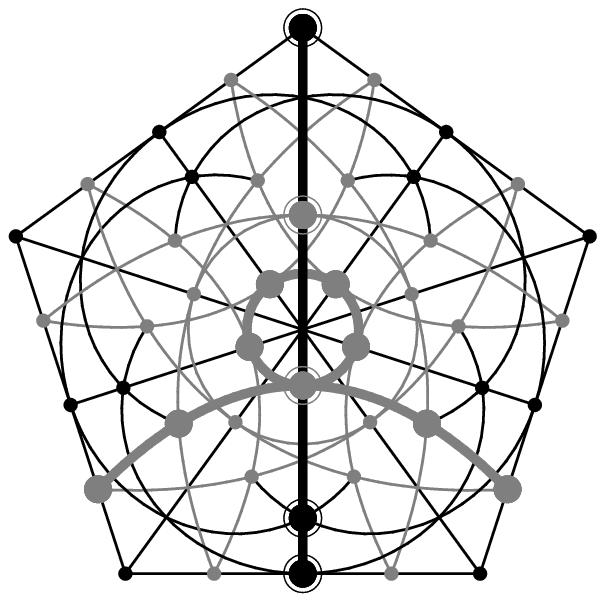}}
    % \put(3.375,0){$A$}
  \end{picture}%
  \hfill
  \begin{picture}(8.4,7.8)%%  8.4 wide and 7.8 high
    \put(0,0){\includegraphics[height=8\unitlength]{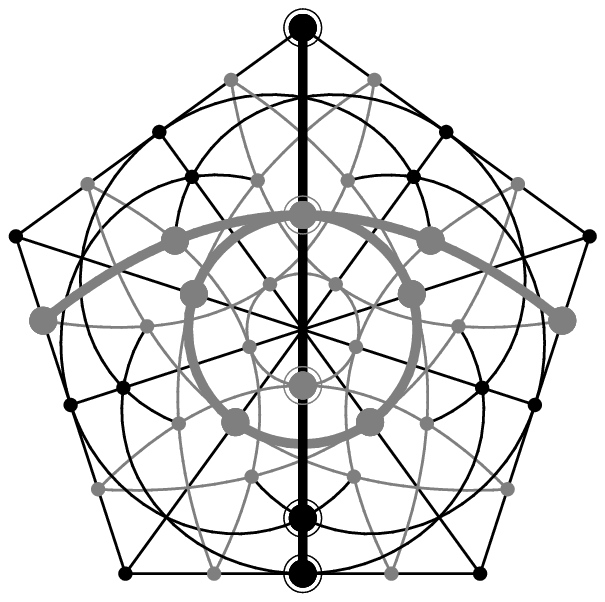}}
    % \put(3.375,0){$A$}
  \end{picture}%
  \hfill
  \begin{picture}(8.4,7.8)%%  8.4 wide and 7.8 high
    \put(0,0){\includegraphics[height=8\unitlength]{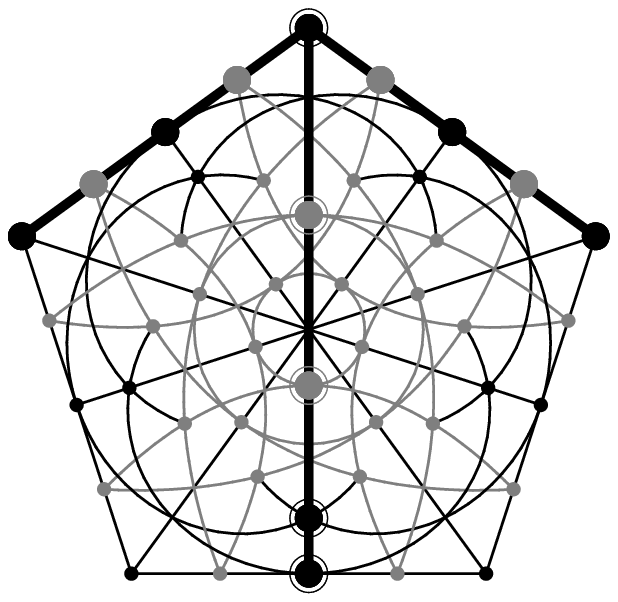}}
    % \put(3.375,0){$A$}
  \end{picture}%
  \newline
  \begin{picture}(8.4,7.8)%%  8.4 wide and 7.8 high
    \put(0,0){\includegraphics[height=8\unitlength]{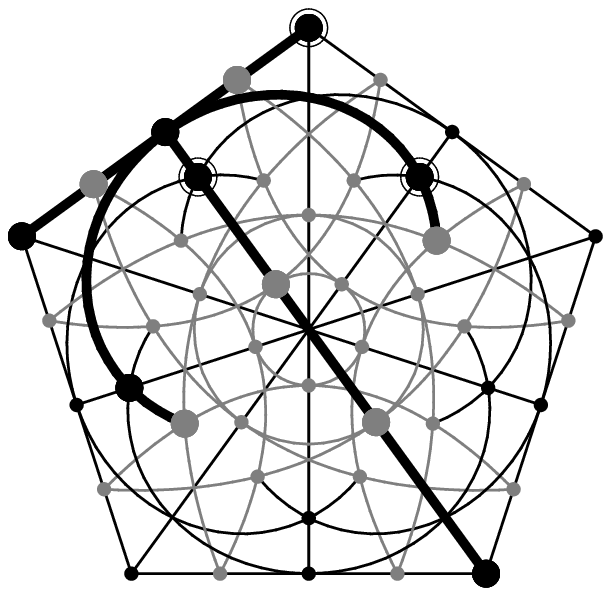}}
    % \put(3.375,0){$A$}
  \end{picture}%
  \hfill
  \begin{picture}(8.4,7.8)%%  8.4 wide and 7.8 high
    \put(0,0){\includegraphics[height=8\unitlength]{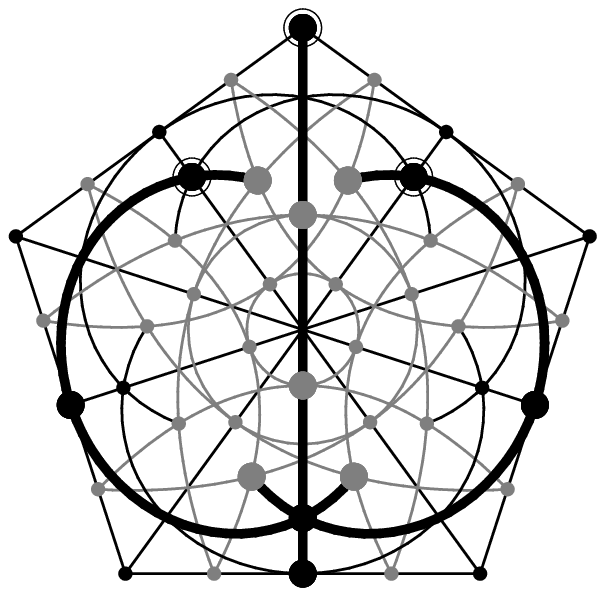}}
    % \put(3.375,0){$A$}
  \end{picture}%
  \hfill
  \begin{picture}(8.4,7.8)%%  8.4 wide and 7.8 high
    \put(0,0){\includegraphics[height=8\unitlength]{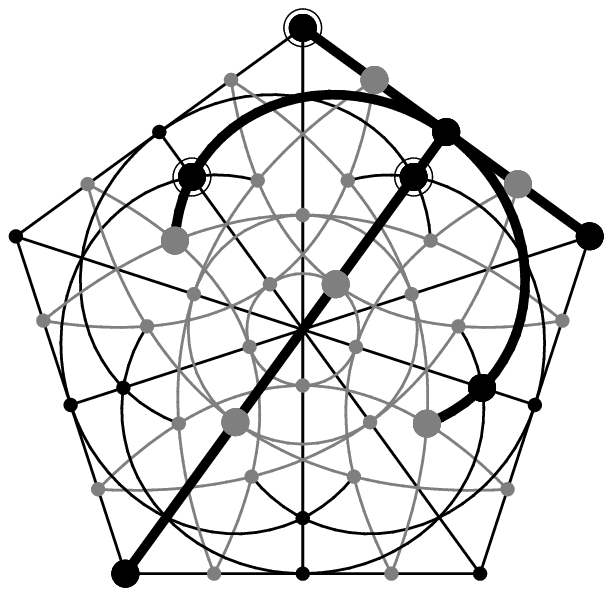}}
    % \put(3.375,0){$A$}
  \end{picture}%
  \hfill
  \begin{picture}(8.4,7.8)%%  8.4 wide and 7.8 high
    \put(0,0){\includegraphics[height=8\unitlength]{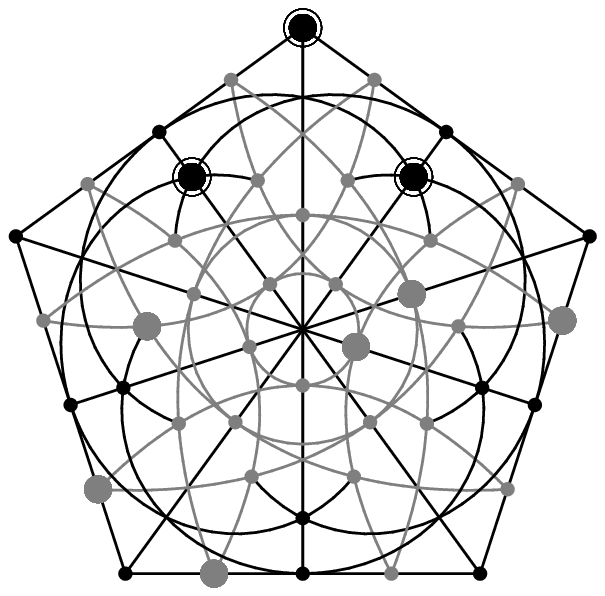}}
    % \put(3.375,0){$A$}
  \end{picture}%
  \hfill
  \begin{picture}(8.4,7.8)%%  8.4 wide and 7.8 high
    \put(0,0){\includegraphics[height=8\unitlength]{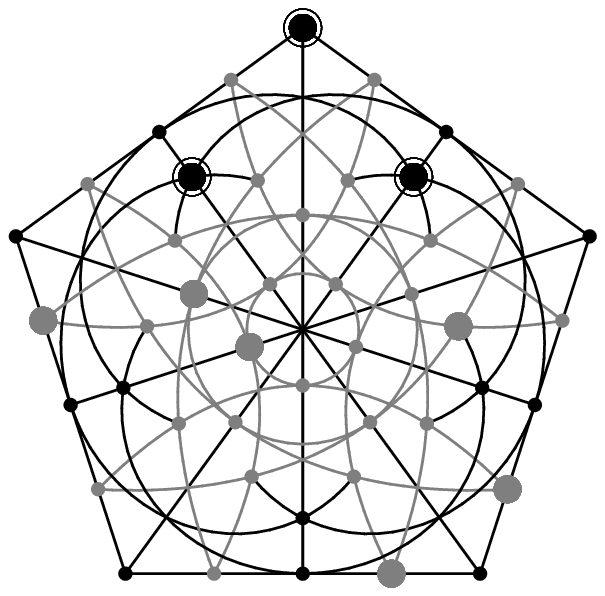}}
    % \put(3.375,0){$A$}
  \end{picture}%
  \newline
  \begin{picture}(8.4,7.8)%%  8.4 wide and 7.8 high
    \put(0,0){\includegraphics[height=8\unitlength]{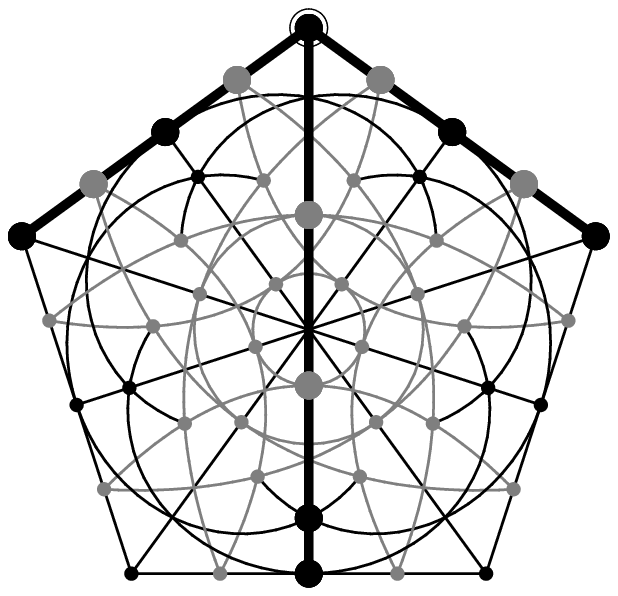}}
    % \put(3.375,0){$A$}
  \end{picture}%
  \hfill
  \begin{picture}(8.4,7.8)%%  8.4 wide and 7.8 high
    \put(0,0){\includegraphics[height=8\unitlength]{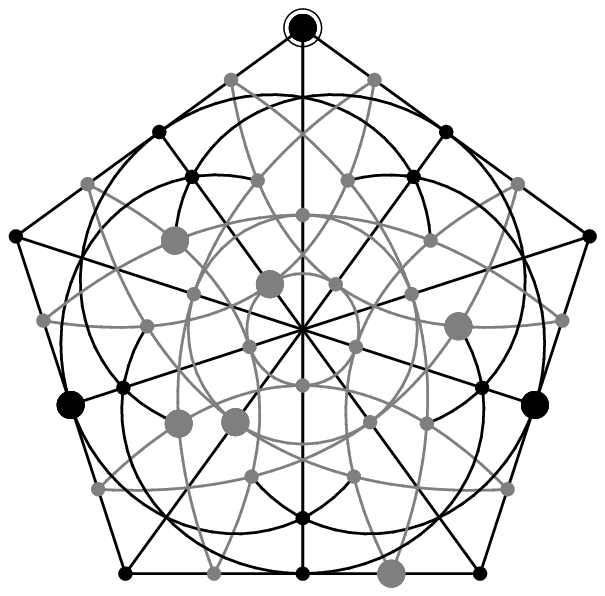}}
    % \put(3.375,0){$A$}
  \end{picture}%
  \hfill
  \begin{picture}(8.4,7.8)%%  8.4 wide and 7.8 high
    \put(0,0){\includegraphics[height=8\unitlength]{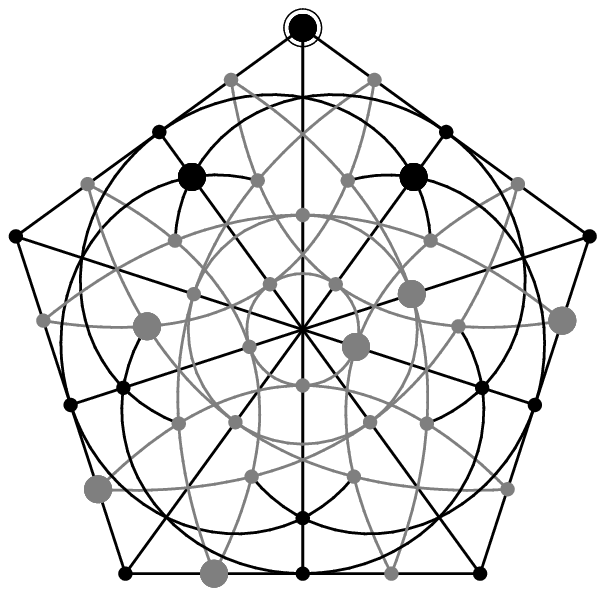}}
    % \put(3.375,0){$A$}
  \end{picture}%
  \hfill
  \begin{picture}(8.4,7.8)%%  8.4 wide and 7.8 high
    \put(0,0){\includegraphics[height=8\unitlength]{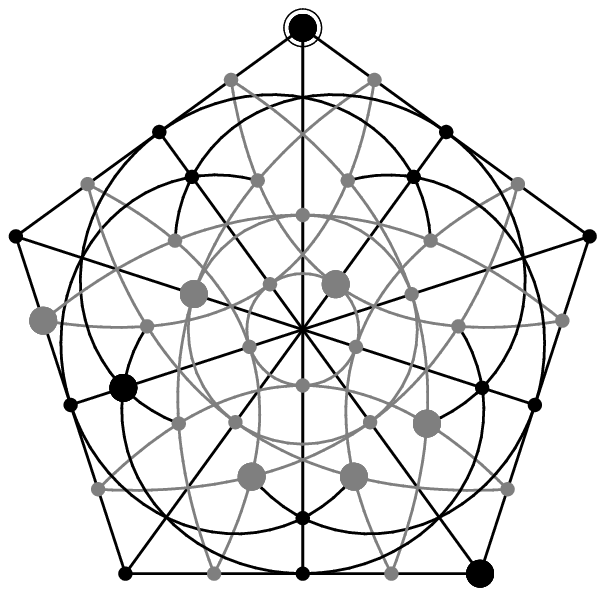}}
    % \put(3.375,0){$A$}
  \end{picture}%
  \hfill
  \begin{picture}(8.4,7.8)%%  8.4 wide and 7.8 high
    \put(0,0){\includegraphics[height=8\unitlength]{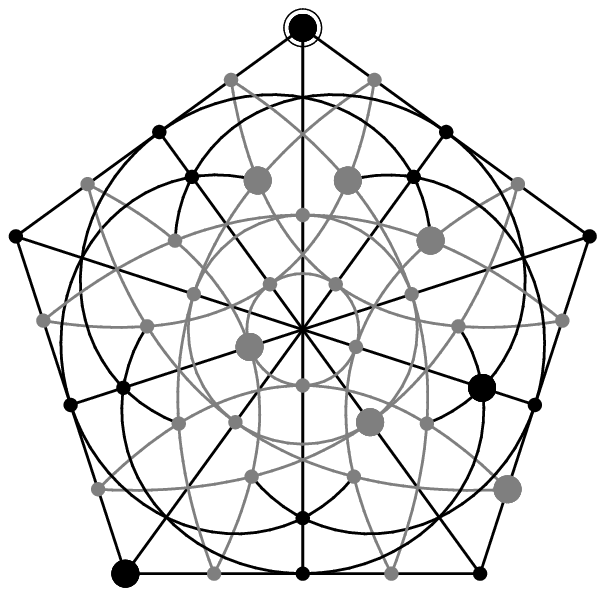}}
    % \put(3.375,0){$A$}
  \end{picture}%
  \newline
  \begin{picture}(8.4,7.8)%%  8.4 wide and 7.8 high
    \put(0,0){\includegraphics[height=8\unitlength]{fig2a.eps}}
    % \put(3.375,0){$A$}
  \end{picture}%
  \hfill
  \begin{picture}(8.4,7.8)%%  8.4 wide and 7.8 high
    \put(0,0){\includegraphics[height=8\unitlength]{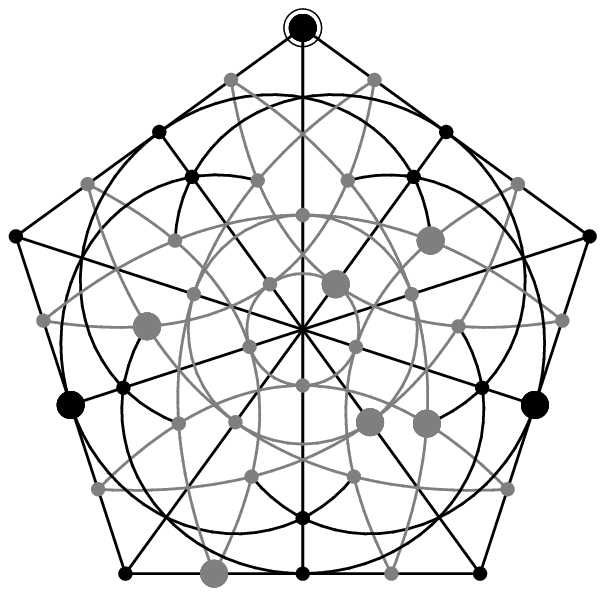}}
    % \put(3.375,0){$A$}
  \end{picture}%
  \hfill
  \begin{picture}(8.4,7.8)%%  8.4 wide and 7.8 high
    \put(0,0){\includegraphics[height=8\unitlength]{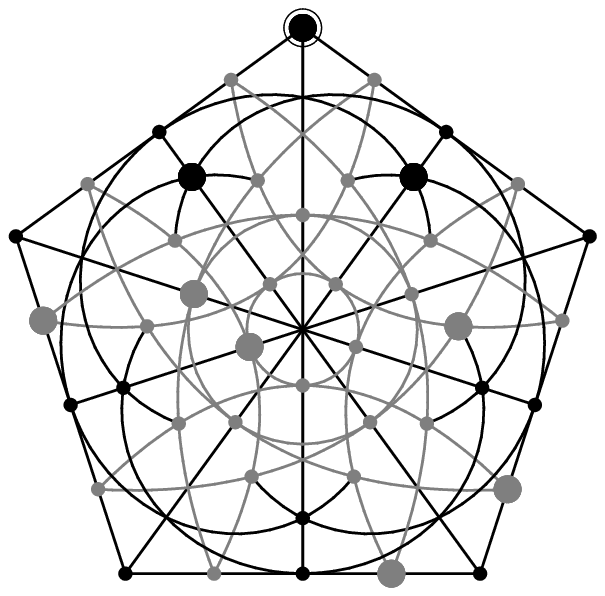}}
    % \put(3.375,0){$A$}
  \end{picture}%
  \hfill
  \begin{picture}(8.4,7.8)%%  8.4 wide and 7.8 high
    \put(0,0){\includegraphics[height=8\unitlength]{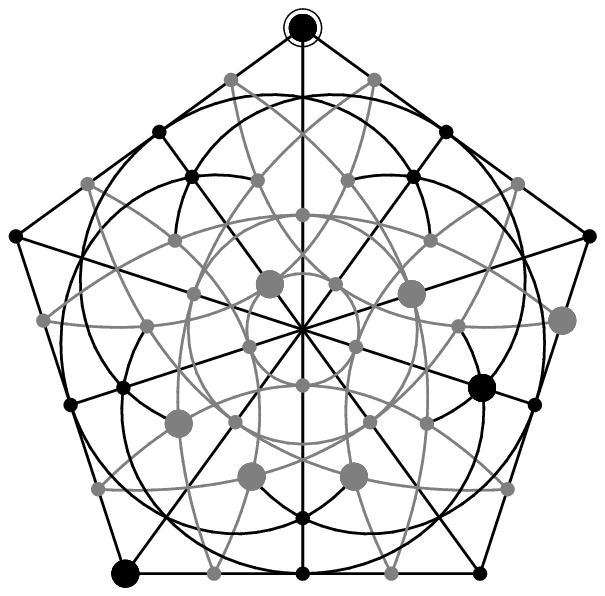}}
    % \put(3.375,0){$A$}
  \end{picture}%
  \hfill
  \begin{picture}(8.4,7.8)%%  8.4 wide and 7.8 high
    \put(0,0){\includegraphics[height=8\unitlength]{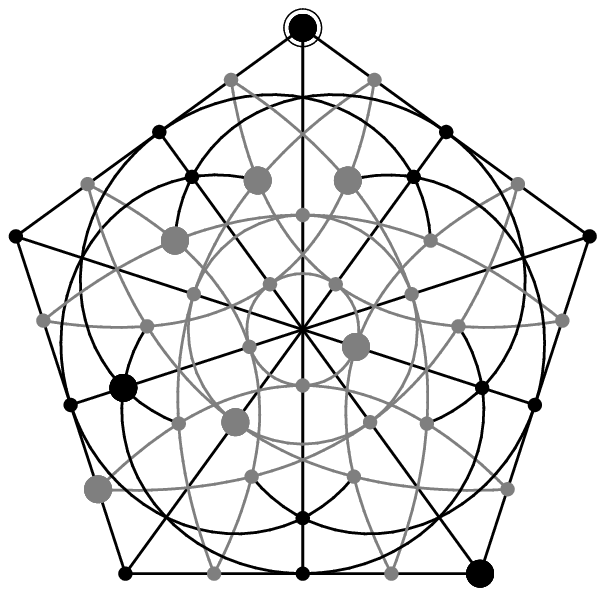}}
    % \put(3.375,0){$A$}
  \end{picture}%
  %\hfill{}
    \caption{Representatives of the four different types of V-lines
    forming the linear subspace of $\mathcal{V}$(GQ(4,\,2) isomorphic to PG(3,\,4).
    The encircled bullets represent the points shared by all the five V-points forming a given V-line.}
\end{figure}

\subsection{Examples of V-lines on Three and Two Common Points}
In order to show that $\mathcal{V}$(GQ(4,\,2)) is not a (partial)
linear space it suffices to find two (or more) V-lines sharing two
(or more) V-points. From the previous subsection it should already
be fairly obvious that it is the existence of tripods that
prevents  $\mathcal{V}$(GQ(4,\,2)) from being a linear space.

To this end, let us have again a look at a type-two V-line of the
PG(3,\,4)-subspace, reproduced once again in Figure 3, top row. It
turns out that we can get a new type of V-line by replacing its
two plane ovoids by two particular tripods, as shown in Figure 3,
bottom row. Hence, we have an example of two distinct V-lines
having three V-points (the three perps) in common.

\begin{figure}[!h]\unitlength0.35cm%%  8.4 breit und 7.8 hoch
\centering
  %{}\hfill
  \begin{picture}(8.4,7.8)%%  8.4 wide and 7.8 high
    \put(0,0){\includegraphics[height=8\unitlength]{fig2j.eps}}
    % \put(3.375,0){$A$}
  \end{picture}%
  \hfill
  \begin{picture}(8.4,7.8)%%  8.4 wide and 7.8 high
    \put(0,0){\includegraphics[height=8\unitlength]{fig2k.eps}}
    % \put(3.375,0){$A$}
  \end{picture}%
  \hfill
  \begin{picture}(8.4,7.8)%%  8.4 wide and 7.8 high
    \put(0,0){\includegraphics[height=8\unitlength]{fig2l.eps}}
    % \put(3.375,0){$A$}
  \end{picture}%
  \hfill
  \begin{picture}(8.4,7.8)%%  8.4 wide and 7.8 high
    \put(0,0){\includegraphics[height=8\unitlength]{fig2m.eps}}
    % \put(3.375,0){$A$}
  \end{picture}%
  \hfill
  \begin{picture}(8.4,7.8)%%  8.4 wide and 7.8 high
    \put(0,0){\includegraphics[height=8\unitlength]{fig2n.eps}}
    % \put(3.375,0){$A$}
  \end{picture}%
  \newline
  \begin{picture}(8.4,7.8)%%  8.4 wide and 7.8 high
    \put(0,0){\includegraphics[height=8\unitlength]{fig2j.eps}}
    % \put(3.375,0){$A$}
  \end{picture}%
  \hfill
  \begin{picture}(8.4,7.8)%%  8.4 wide and 7.8 high
    \put(0,0){\includegraphics[height=8\unitlength]{fig2k.eps}}
    % \put(3.375,0){$A$}
  \end{picture}%
  \hfill
  \begin{picture}(8.4,7.8)%%  8.4 wide and 7.8 high
    \put(0,0){\includegraphics[height=8\unitlength]{fig2l.eps}}
    % \put(3.375,0){$A$}
  \end{picture}%
  \hfill
  \begin{picture}(8.4,7.8)%%  8.4 wide and 7.8 high
    \put(0,0){\includegraphics[height=8\unitlength]{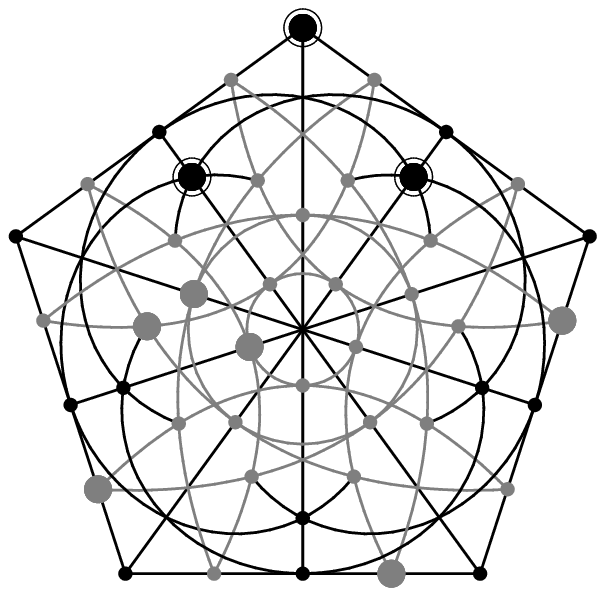}}
    % \put(3.375,0){$A$}
  \end{picture}%
  \hfill
  \begin{picture}(8.4,7.8)%%  8.4 wide and 7.8 high
    \put(0,0){\includegraphics[height=8\unitlength]{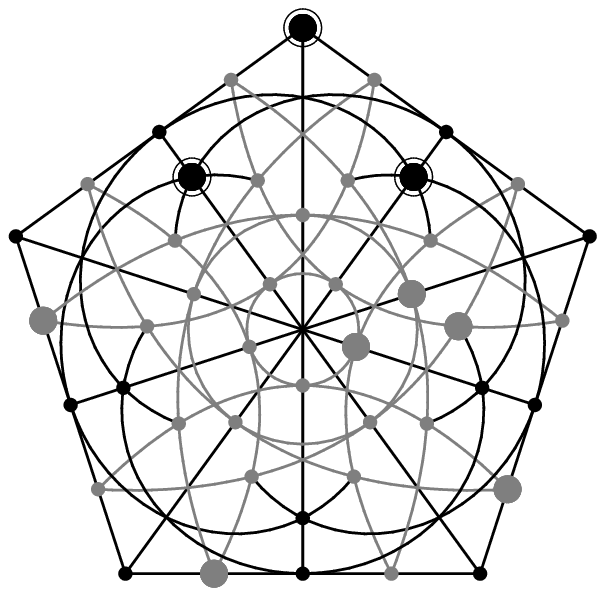}}
    % \put(3.375,0){$A$}
  \end{picture}%
  %\hfill{}
    \caption{An instance of two distinct V-lines having three V-points in common.
    We note in passing that the tripods are tri-tripods with respect to the selected copy of GQ(2,\,2).}
\end{figure}

The next couple of examples feature three (Figure 4) and four
(Figure 5) V-lines through two common V-points. In the first case,
the V-lines consist each of a perp and four tripods in the {\it
rosette} centered at the perp's center; the perp and the first
tripod are the common V-points. With respect to the selected
GQ(2,\,2), the first two tripods in each V-line are tri-tripods,
the other two being uni-tripods. In the second case, each of the
four V-lines represents a {\it fan} of ovoids, the first ovoid
being planar, second a tri-tripod, and the remaining three
uni-tripods.

\begin{figure}[!h]\unitlength0.35cm%%  8.4 breit und 7.8 hoch
\centering
  %{}\hfill
  \begin{picture}(8.4,7.8)%%  8.4 wide and 7.8 high
    \put(0,0){\includegraphics[height=8\unitlength]{fig2a.eps}}
    % \put(3.375,0){$A$}
  \end{picture}%
  \hfill
  \begin{picture}(8.4,7.8)%%  8.4 wide and 7.8 high
    \put(0,0){\includegraphics[height=8\unitlength]{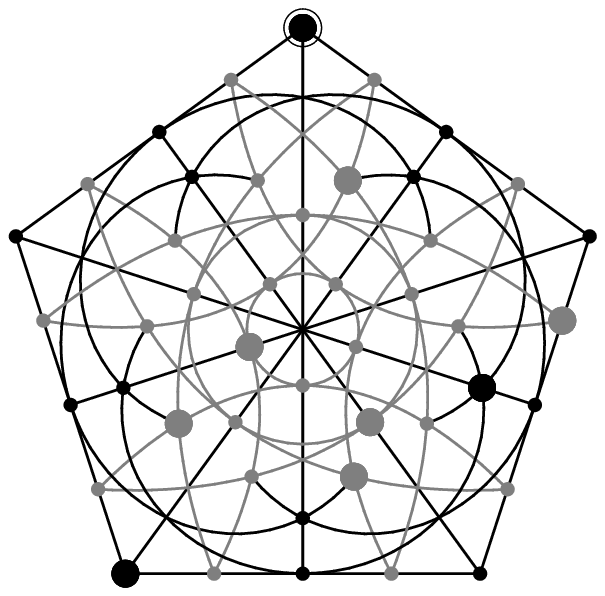}}
    % \put(3.375,0){$A$}
  \end{picture}%
  \hfill
  \begin{picture}(8.4,7.8)%%  8.4 wide and 7.8 high
    \put(0,0){\includegraphics[height=8\unitlength]{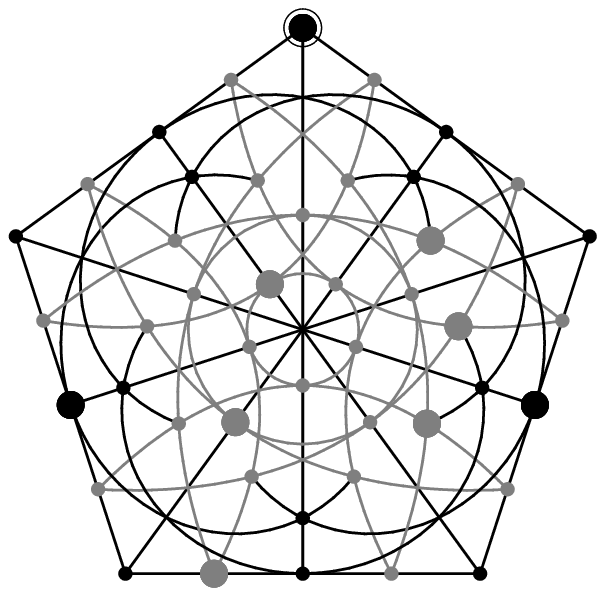}}
    % \put(3.375,0){$A$}
  \end{picture}%
  \hfill
  \begin{picture}(8.4,7.8)%%  8.4 wide and 7.8 high
    \put(0,0){\includegraphics[height=8\unitlength]{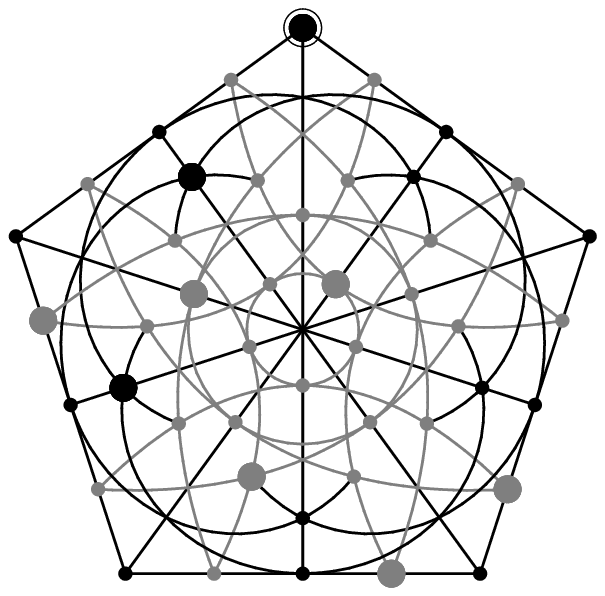}}
    % \put(3.375,0){$A$}
  \end{picture}%
  \hfill
  \begin{picture}(8.4,7.8)%%  8.4 wide and 7.8 high
    \put(0,0){\includegraphics[height=8\unitlength]{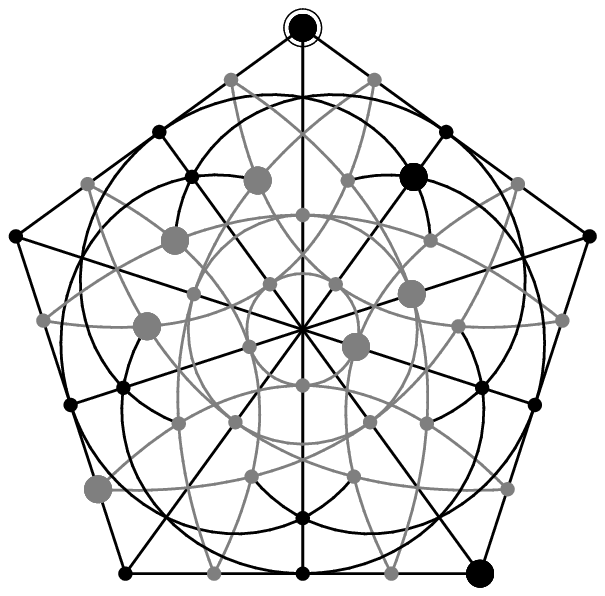}}
    % \put(3.375,0){$A$}
  \end{picture}%
  \newline
  \begin{picture}(8.4,7.8)%%  8.4 wide and 7.8 high
    \put(0,0){\includegraphics[height=8\unitlength]{fig2a.eps}}
    % \put(3.375,0){$A$}
  \end{picture}%
  \hfill
  \begin{picture}(8.4,7.8)%%  8.4 wide and 7.8 high
    \put(0,0){\includegraphics[height=8\unitlength]{fig4b.eps}}
    % \put(3.375,0){$A$}
  \end{picture}%
  \hfill
  \begin{picture}(8.4,7.8)%%  8.4 wide and 7.8 high
    \put(0,0){\includegraphics[height=8\unitlength]{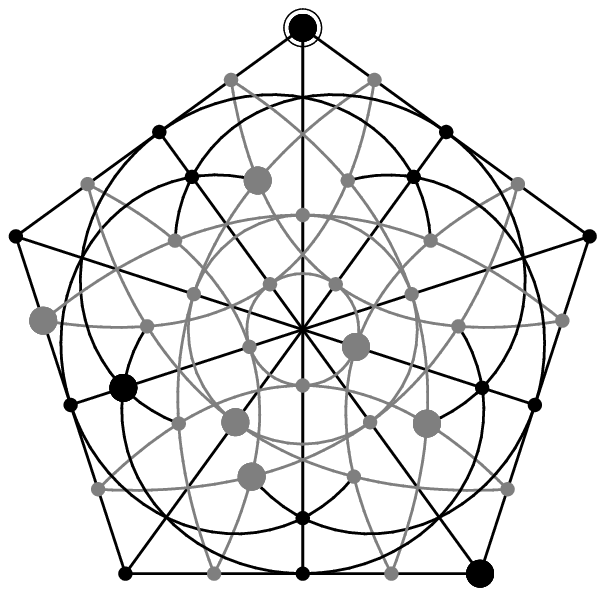}}
    % \put(3.375,0){$A$}
  \end{picture}%
  \hfill
  \begin{picture}(8.4,7.8)%%  8.4 wide and 7.8 high
    \put(0,0){\includegraphics[height=8\unitlength]{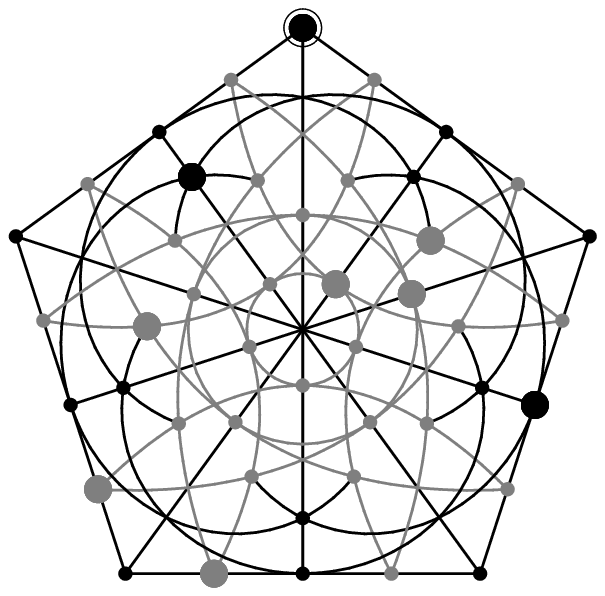}}
    % \put(3.375,0){$A$}
  \end{picture}%
  \hfill
  \begin{picture}(8.4,7.8)%%  8.4 wide and 7.8 high
    \put(0,0){\includegraphics[height=8\unitlength]{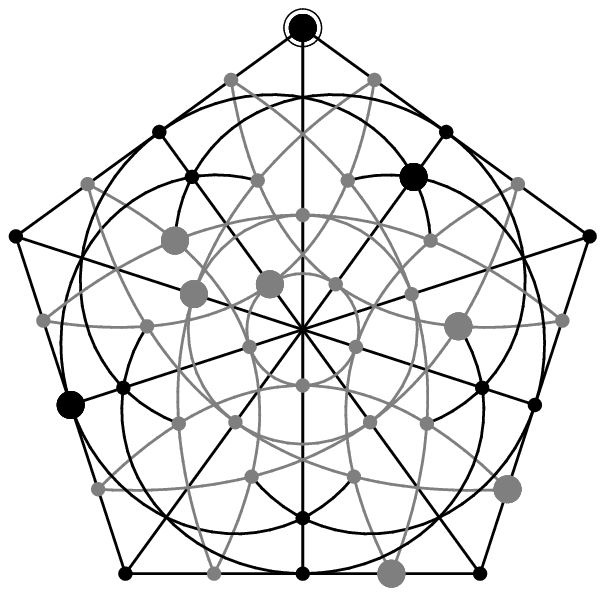}}
    % \put(3.375,0){$A$}
  \end{picture}%
  \newline
  \begin{picture}(8.4,7.8)%%  8.4 wide and 7.8 high
    \put(0,0){\includegraphics[height=8\unitlength]{fig2a.eps}}
    % \put(3.375,0){$A$}
  \end{picture}%
  \hfill
  \begin{picture}(8.4,7.8)%%  8.4 wide and 7.8 high
    \put(0,0){\includegraphics[height=8\unitlength]{fig4b.eps}}
    % \put(3.375,0){$A$}
  \end{picture}%
  \hfill
  \begin{picture}(8.4,7.8)%%  8.4 wide and 7.8 high
    \put(0,0){\includegraphics[height=8\unitlength]{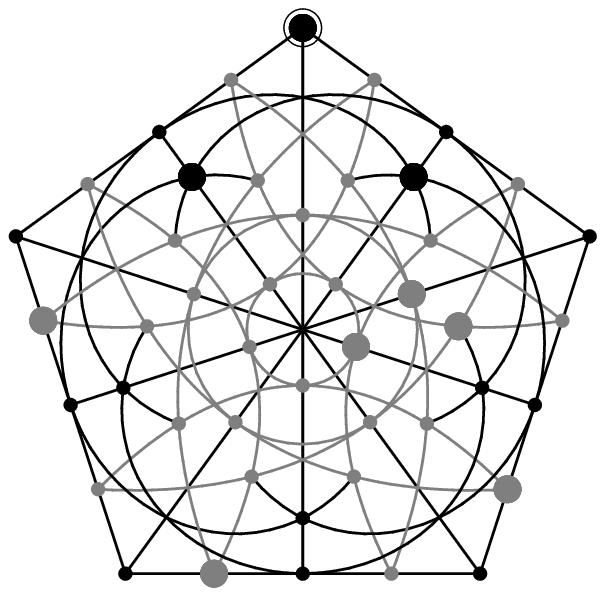}}
    % \put(3.375,0){$A$}
  \end{picture}%
  \hfill
  \begin{picture}(8.4,7.8)%%  8.4 wide and 7.8 high
    \put(0,0){\includegraphics[height=8\unitlength]{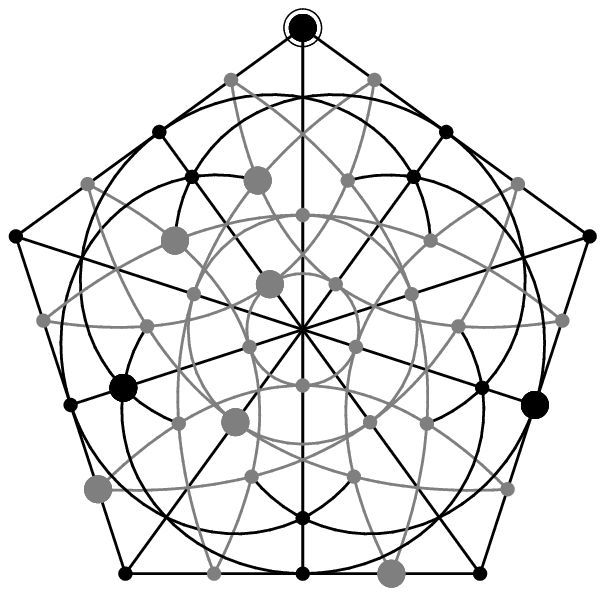}}
    % \put(3.375,0){$A$}
  \end{picture}%
  \hfill
  \begin{picture}(8.4,7.8)%%  8.4 wide and 7.8 high
    \put(0,0){\includegraphics[height=8\unitlength]{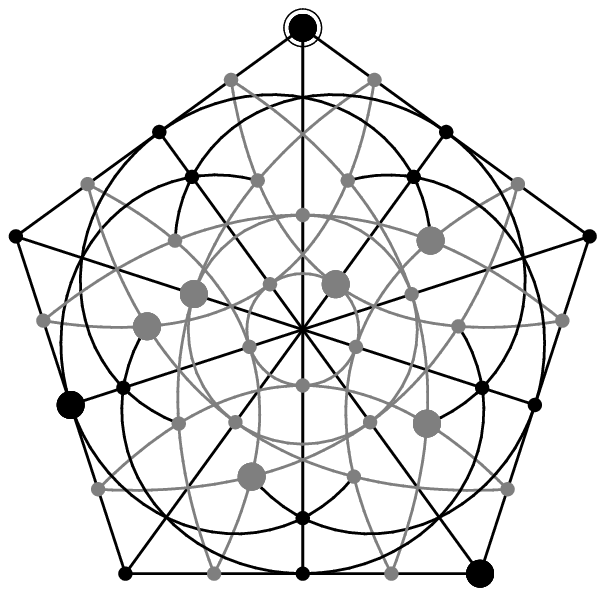}}
    % \put(3.375,0){$A$}
  \end{picture}%
  %\hfill{}
    \caption{An example of three different V-lines on two common V-points (a perp and a (tri-)tripod).}
\end{figure}

\begin{figure}[!h]\unitlength0.35cm%%  8.4 breit und 7.8 hoch
\centering
  %{}\hfill
  \begin{picture}(8.4,7.8)%%  8.4 wide and 7.8 high
    \put(0,0){\includegraphics[height=8\unitlength]{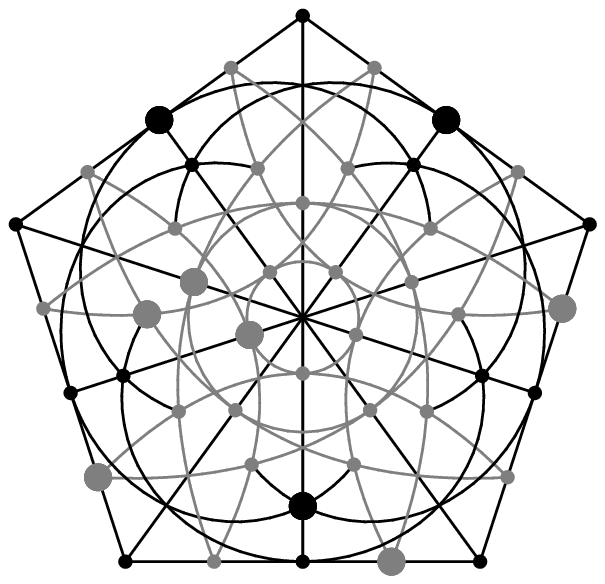}}
    % \put(3.375,0){$A$}
  \end{picture}%
  \hfill
  \begin{picture}(8.4,7.8)%%  8.4 wide and 7.8 high
    \put(0,0){\includegraphics[height=8\unitlength]{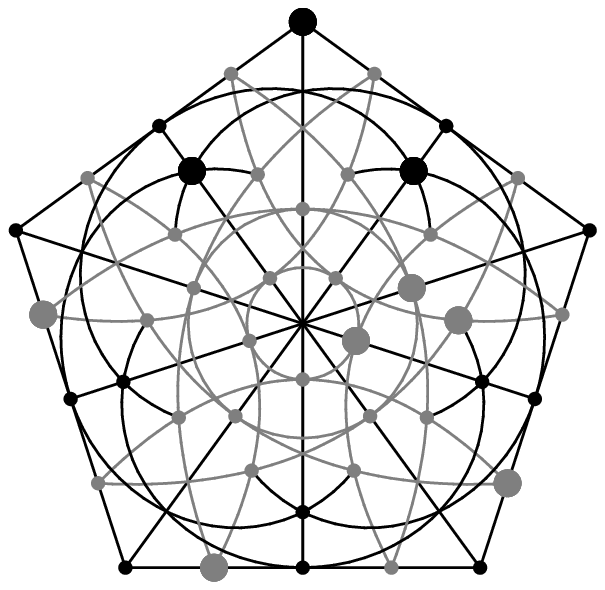}}
    % \put(3.375,0){$A$}
  \end{picture}%
  \hfill
  \begin{picture}(8.4,7.8)%%  8.4 wide and 7.8 high
    \put(0,0){\includegraphics[height=8\unitlength]{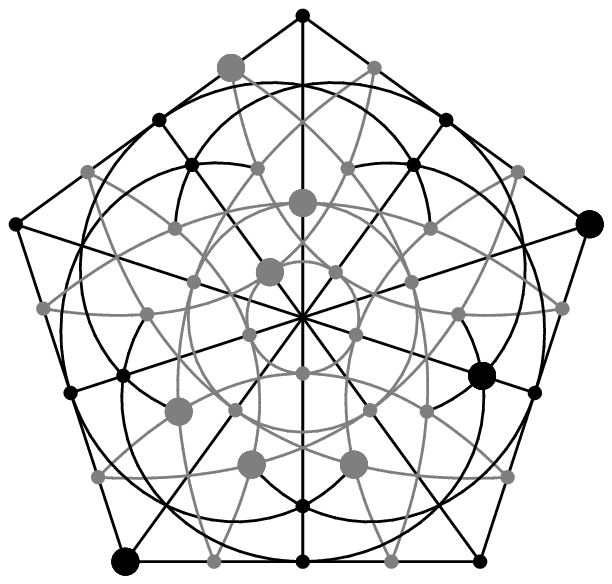}}
    % \put(3.375,0){$A$}
  \end{picture}%
  \hfill
  \begin{picture}(8.4,7.8)%%  8.4 wide and 7.8 high
    \put(0,0){\includegraphics[height=8\unitlength]{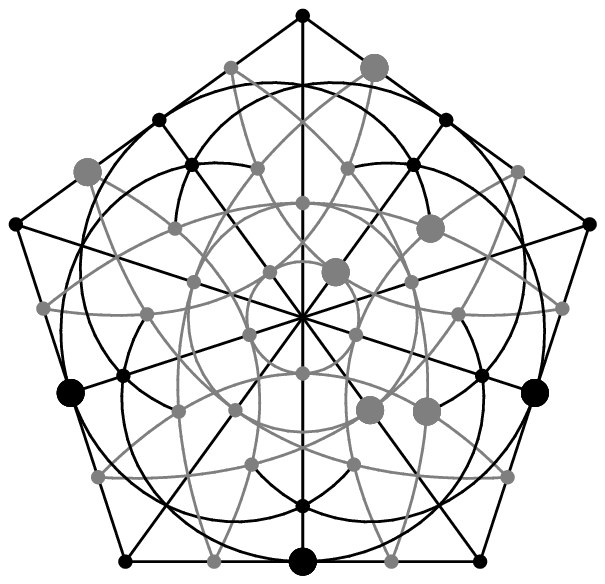}}
    % \put(3.375,0){$A$}
  \end{picture}%
  \hfill
  \begin{picture}(8.4,7.8)%%  8.4 wide and 7.8 high
    \put(0,0){\includegraphics[height=8\unitlength]{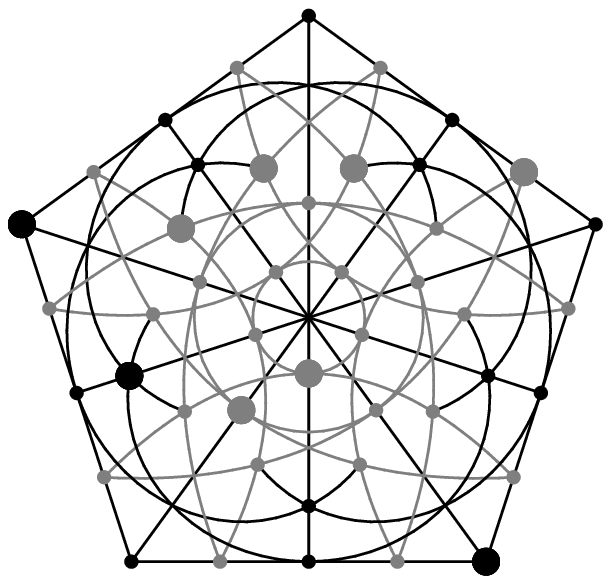}}
    % \put(3.375,0){$A$}
  \end{picture}%
  \newline
  \begin{picture}(8.4,7.8)%%  8.4 wide and 7.8 high
    \put(0,0){\includegraphics[height=8\unitlength]{fig5a.eps}}
    % \put(3.375,0){$A$}
  \end{picture}%
  \hfill
  \begin{picture}(8.4,7.8)%%  8.4 wide and 7.8 high
    \put(0,0){\includegraphics[height=8\unitlength]{fig5b.eps}}
    % \put(3.375,0){$A$}
  \end{picture}%
  \hfill
  \begin{picture}(8.4,7.8)%%  8.4 wide and 7.8 high
    \put(0,0){\includegraphics[height=8\unitlength]{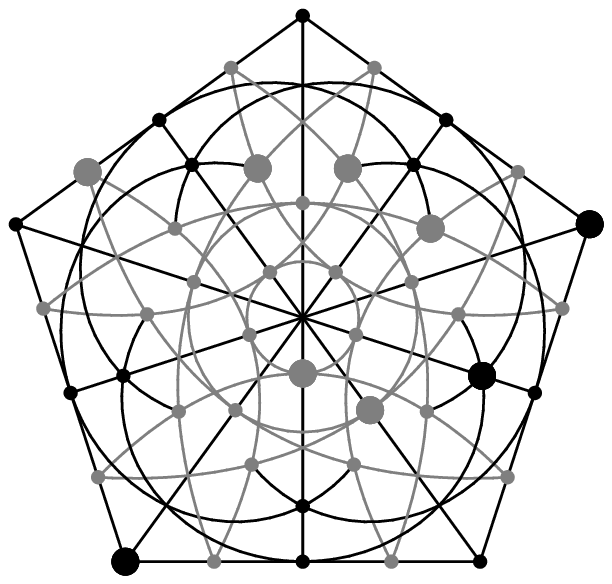}}
    % \put(3.375,0){$A$}
  \end{picture}%
  \hfill
  \begin{picture}(8.4,7.8)%%  8.4 wide and 7.8 high
    \put(0,0){\includegraphics[height=8\unitlength]{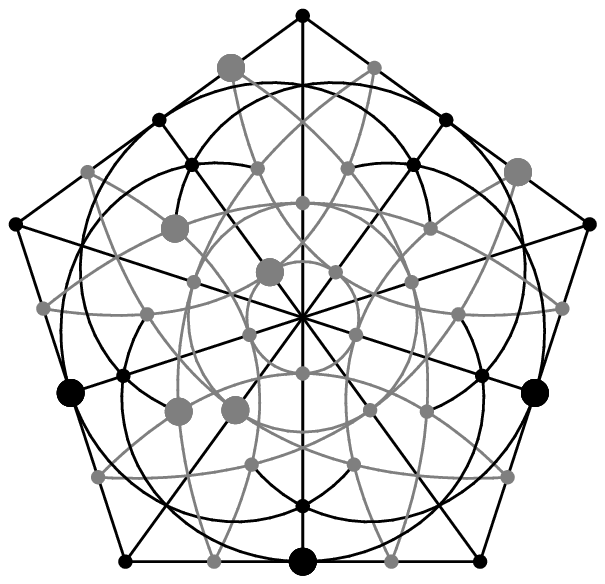}}
    % \put(3.375,0){$A$}
  \end{picture}%
  \hfill
  \begin{picture}(8.4,7.8)%%  8.4 wide and 7.8 high
    \put(0,0){\includegraphics[height=8\unitlength]{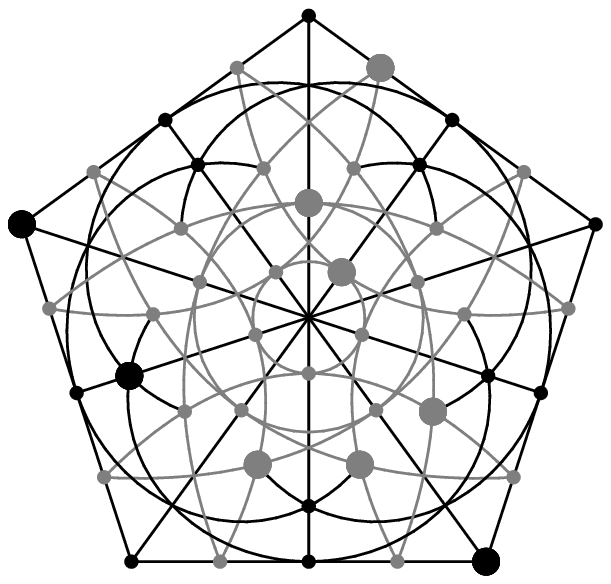}}
    % \put(3.375,0){$A$}
  \end{picture}%
  \newline
  \begin{picture}(8.4,7.8)%%  8.4 wide and 7.8 high
    \put(0,0){\includegraphics[height=8\unitlength]{fig5a.eps}}
    % \put(3.375,0){$A$}
  \end{picture}%
  \hfill
  \begin{picture}(8.4,7.8)%%  8.4 wide and 7.8 high
    \put(0,0){\includegraphics[height=8\unitlength]{fig5b.eps}}
    % \put(3.375,0){$A$}
  \end{picture}%
  \hfill
  \begin{picture}(8.4,7.8)%%  8.4 wide and 7.8 high
    \put(0,0){\includegraphics[height=8\unitlength]{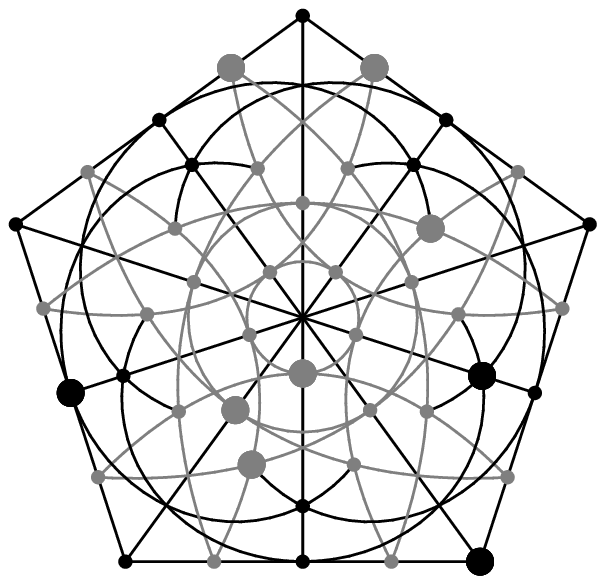}}
    % \put(3.375,0){$A$}
  \end{picture}%
  \hfill
  \begin{picture}(8.4,7.8)%%  8.4 wide and 7.8 high
    \put(0,0){\includegraphics[height=8\unitlength]{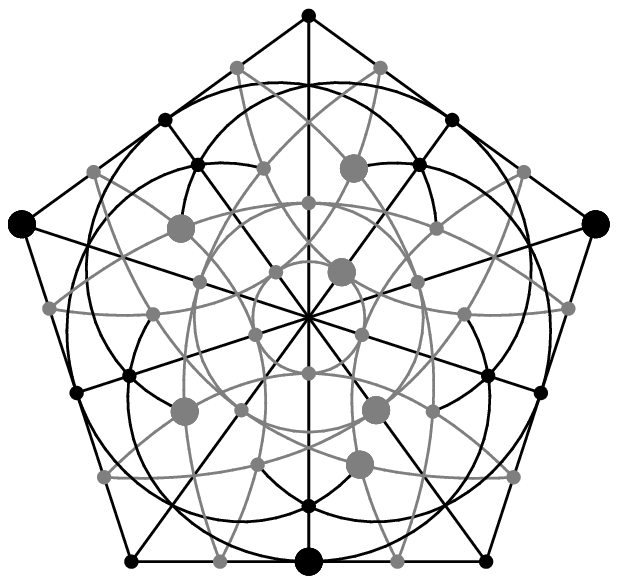}}
    % \put(3.375,0){$A$}
  \end{picture}%
  \hfill
  \begin{picture}(8.4,7.8)%%  8.4 wide and 7.8 high
    \put(0,0){\includegraphics[height=8\unitlength]{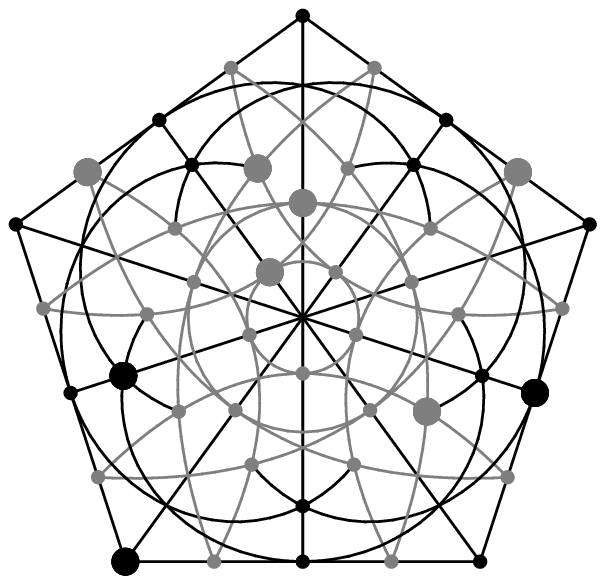}}
    % \put(3.375,0){$A$}
  \end{picture}%
  \newline
  \begin{picture}(8.4,7.8)%%  8.4 wide and 7.8 high
    \put(0,0){\includegraphics[height=8\unitlength]{fig5a.eps}}
    % \put(3.375,0){$A$}
  \end{picture}%
  \hfill
  \begin{picture}(8.4,7.8)%%  8.4 wide and 7.8 high
    \put(0,0){\includegraphics[height=8\unitlength]{fig5b.eps}}
    % \put(3.375,0){$A$}
  \end{picture}%
  \hfill
  \begin{picture}(8.4,7.8)%%  8.4 wide and 7.8 high
    \put(0,0){\includegraphics[height=8\unitlength]{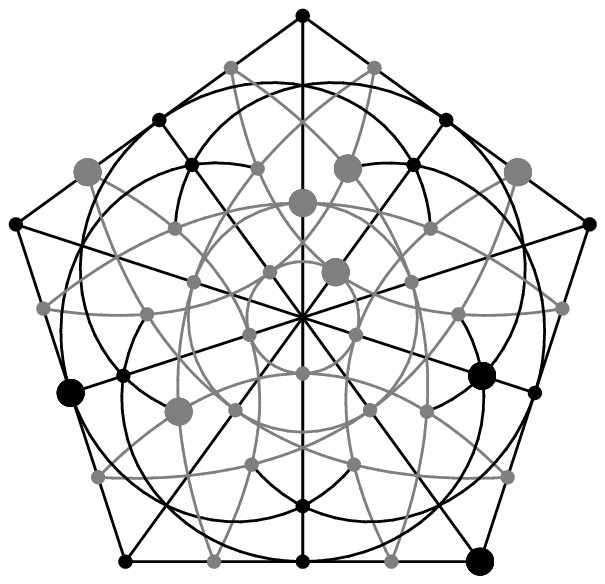}}
    % \put(3.375,0){$A$}
  \end{picture}%
  \hfill
  \begin{picture}(8.4,7.8)%%  8.4 wide and 7.8 high
    \put(0,0){\includegraphics[height=8\unitlength]{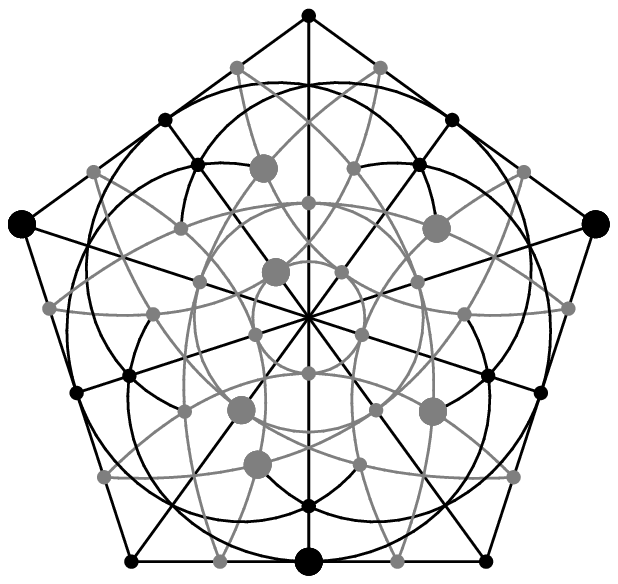}}
    % \put(3.375,0){$A$}
  \end{picture}%
  \hfill
  \begin{picture}(8.4,7.8)%%  8.4 wide and 7.8 high
    \put(0,0){\includegraphics[height=8\unitlength]{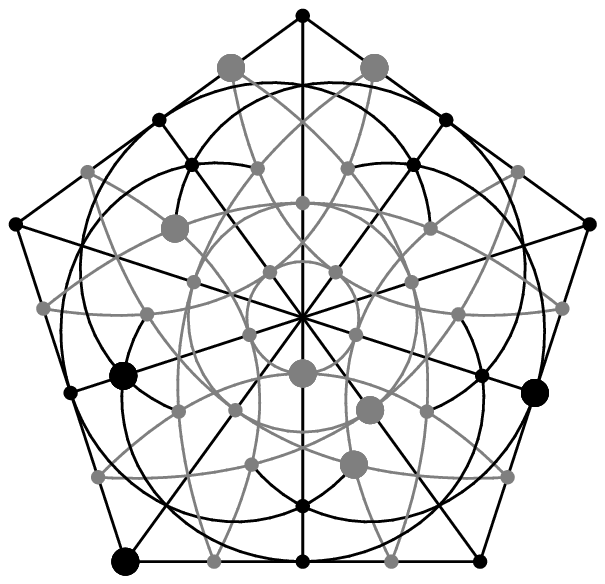}}
    % \put(3.375,0){$A$}
  \end{picture}%
 % \hfill{}
    \caption{An example of four different V-lines sharing two V-points (a plane ovoid and a (tri-)tripod).}
\end{figure}

\subsection{Examples of V-lines of Size Three and Two}
That the structure of $\mathcal{V}$(GQ(4,\,2)) is much more
complex and intricate than that of any (partial) linear space is,
alongside the above-introduced examples, also illustrated by the
existence of V-lines of cardinality less than five. Thus, we found
many V-lines of size three, like the one shown in Figure 6. This
V-line consists of a perp and two (uni-)tripods on a common
unicentric triad; since two perps, obviously, cannot share a
unicentric triad and a given perp contains 48 such triads, we find
altogether $45 \times 48 = 2160$ V-lines of this particular kind.
Finally, there are also a large number of V-lines of size two; the
one depicted in Figure 7 is composed of a plane ovoid and a tripod
having six points in common.

\begin{figure}[!h]\unitlength0.4cm%%  8.4 breit und 7.8 hoch
\centering
  {}\hfill
  \begin{picture}(8.4,7.8)%%  8.4 wide and 7.8 high
    \put(0,0){\includegraphics[height=8\unitlength]{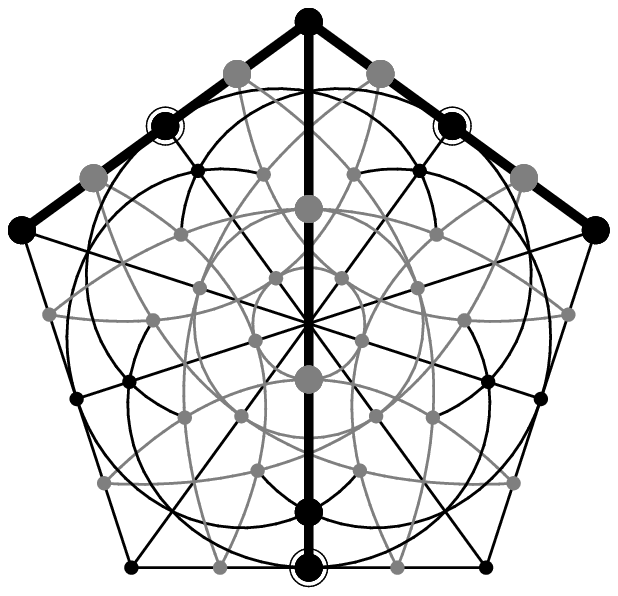}}
    % \put(3.375,0){$A$}
  \end{picture}%
  \hfill
  \begin{picture}(8.4,7.8)%%  8.4 wide and 7.8 high
    \put(0,0){\includegraphics[height=8\unitlength]{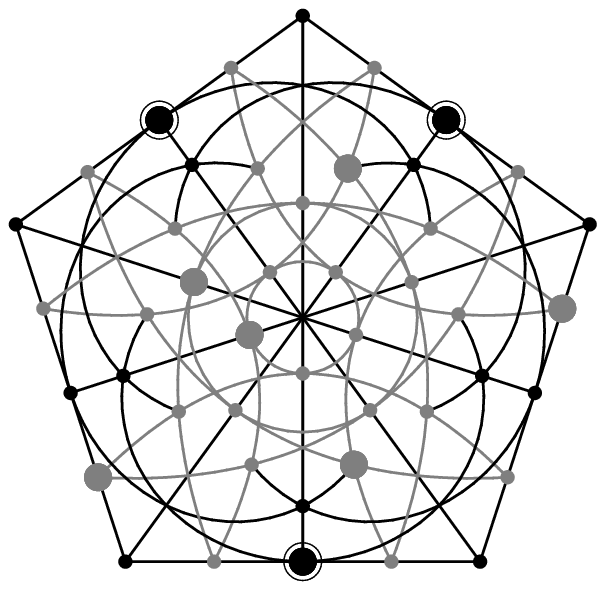}}
    % \put(3.375,0){$A$}
  \end{picture}%
  \hfill
  \begin{picture}(8.4,7.8)%%  8.4 wide and 7.8 high
    \put(0,0){\includegraphics[height=8\unitlength]{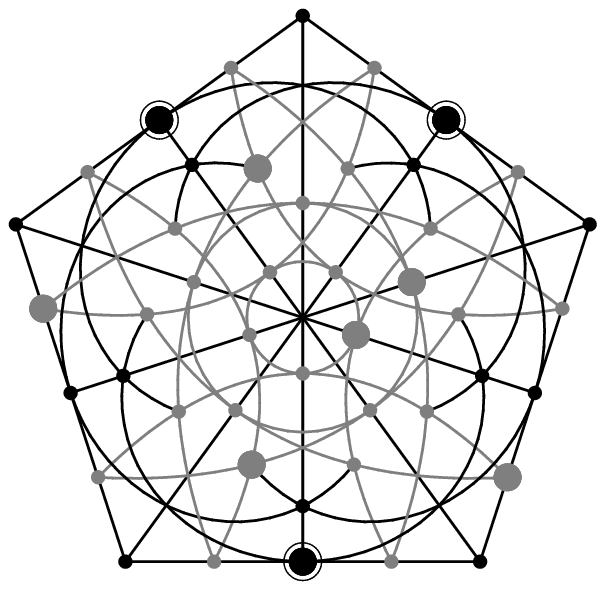}}
    % \put(3.375,0){$A$}
  \end{picture}%
  \hfill{}
    \caption{An example of V-line of size three.}
\end{figure}

\begin{figure}[!h]\unitlength0.4cm%%  8.4 breit und 7.8 hoch
\centering
  {}\hfill
  \begin{picture}(8.4,7.8)%%  8.4 wide and 7.8 high
    \put(0,0){\includegraphics[height=8\unitlength]{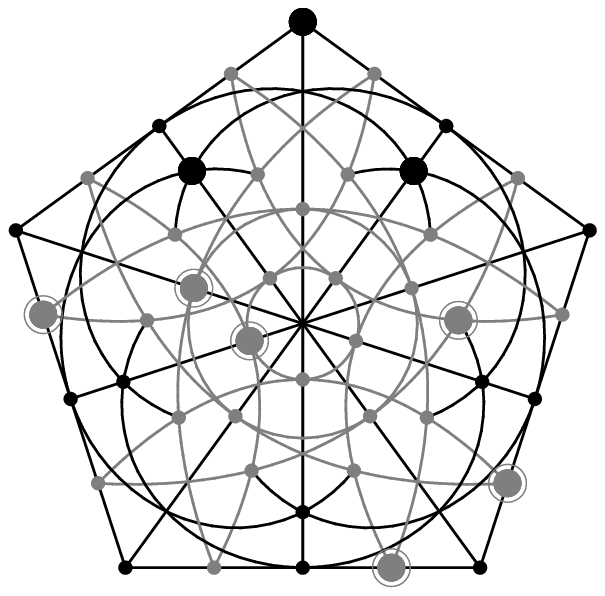}}
    % \put(3.375,0){$A$}
  \end{picture}%
  \hfill
  \begin{picture}(8.4,7.8)%%  8.4 wide and 7.8 high
    \put(0,0){\includegraphics[height=8\unitlength]{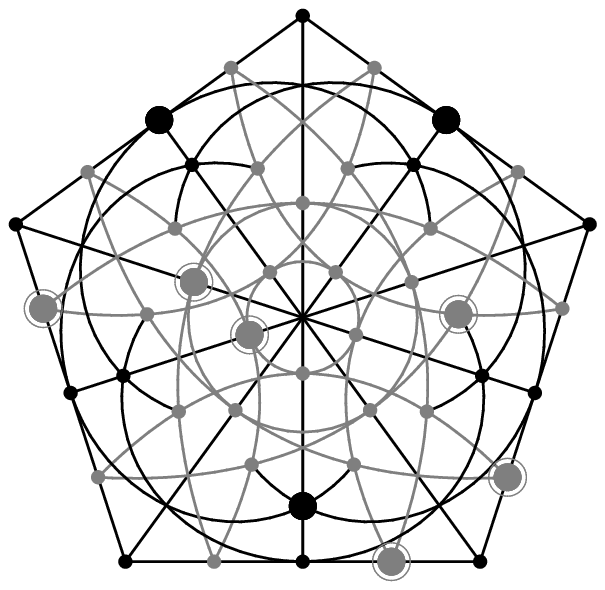}}
    % \put(3.375,0){$A$}
  \end{picture}%
  \hfill{}
    \caption{An example of V-line of size two.}
\end{figure}

\subsection{V-lines as Fans of Ovoids Through a Given Ovoid}

 We shall finish the paper by having a closer look at very impressive subconfigurations of V-lines represented by fans of ovoids.  One first recalls (Sec.\,3) that a fan of ovoids of GQ(4,\,2) is a pentad of (pairwise disjoint) ovoids partitioning the point set. Every fan comprises --- as already shown in Figure 5 ---  a plane ovoid, a tri-tripod and three uni-tripods; in the figures below these are represented by red-, green- and yellow-coloured circles, respectively.

Given a {\it plane} ovoid, there are no plane ovoids, 10 tri-tripods and 30 uni-tripods disjoint from it. Figure 8 (``spider") depicts the configuration of 13 fans of ovoids through a plane ovoid (P). We see a remarkable pattern. One tri-tripod (4) has a special standing as there are four distinct fans passing through it. Similarly, three out of 30 uni-tripods (namely 19, 25 and 28) have a particular footing as there are four distinct fans through each of them. Leaving the distinguished fan $\{$P, 4, 19, 25, 28$\}$ aside, the remaining twelve fans form four classes of cardinality three each. In each class the fans have two ovoids in common (namely P and 4, P and 19, P and 25, and P and 28); whereas in the first class the two ovoids are both plane ovoids, in the remaining three they are of two distinct kinds. Hence, considering the multiplicities and character of mutual  intersection of fans, the totality of the latter is split into five subsets in a [3 + 3 + 3] + (3) + 1 fashion; the ``1" corresponds, naturally, to the distinguished fan $\{$P, 4, 19, 25, 28$\}$.

Given a {\it tri}-tripod, there are 10 plane ovoids, no tri-tripods and 21 uni-tripods disjoint from it. Figure 9 (``bee") depicts the configuration  of 13 fans of ovoids through a tri-tripod (X). We again see a remarkable pattern. One plane ovoid (5) has a special standing as there are four distinct fans passing through it. Similarly, three out of 21 uni-tripods (11, 13 and 20) have a particular footing as there are seven distinct fans through each of them. The split of the whole family of fans again shows a [3 + 3 + 3] + (3) + 1 pattern; the ``1" here symbolizes the distinguished fan $\{$5, X, 11, 13, 20$\}$. This case differs from the previous one in two crucial aspects. First, the three size-three classes here are such that their fans have {\it three} ovoids in common (namely X, 11 and 13;  X, 11 and 20; and X, 13 and 20), whilst the fans of the fourth class share only a couple of ovoids (X and 5). Second, one finds  three pairs of ovoids, namely X and 11, X and 13, and X and 20, such that there go as many as {\it seven} distinct fans through each of them.

\begin{figure}[pht!]
\centerline{\includegraphics[width=9cm,clip=]{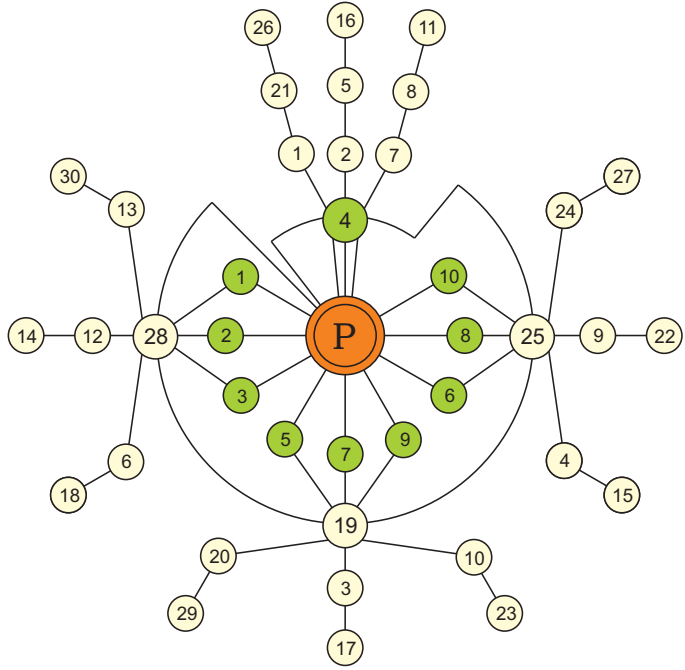}}
\vspace*{.2cm} \caption{A diagrammatic illustration of the qualitative relation between the 13 fans of ovoids sharing a plane ovoid (P). Here and in the following two figures as well, each ovoid is represented  by a numbered circle --- the particulars of the numbering being completely irrelevant for our purpose --- and every fan comprises five different circles (one red, one green and three yellow ones, their order being also irrelevant) joined by broken line-segments and/or arcs of big circles (e.\,g., $\{$P, 1, 28, 13, 30 $\}$,  $\{$P, 2, 28, 12, 14$\}$, $\{$P, 3, 28, 6, 18$\}$, etc.).}
\end{figure}

\begin{figure}[pht!]
\centerline{\includegraphics[width=8cm,clip=]{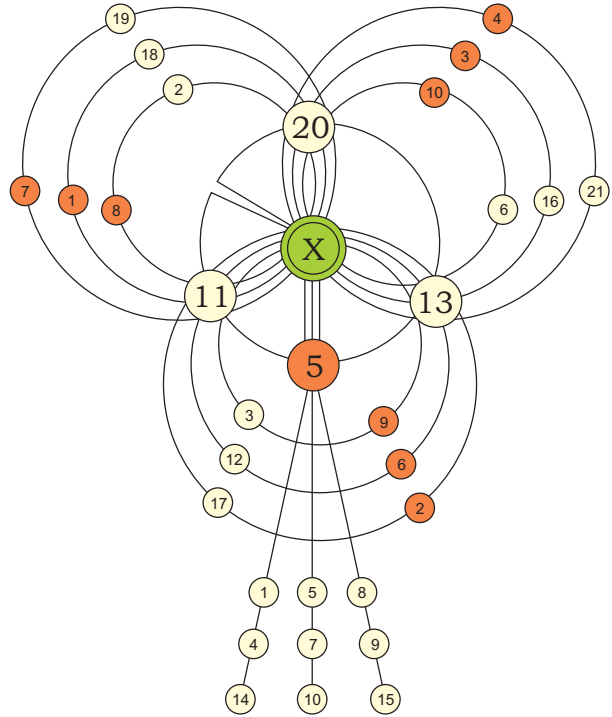}}
\vspace*{.2cm} \caption{The same as in the previous figure for a tri-tripod (X). Note a ``tighter coupling" (i.\,e., higher-multiplicity intersections) between the fans within each of the three classes and between the classes themselves when compared with the plane ovoid case.}
\end{figure}

\begin{figure}[pht!]
\centerline{\includegraphics[width=8cm,clip=]{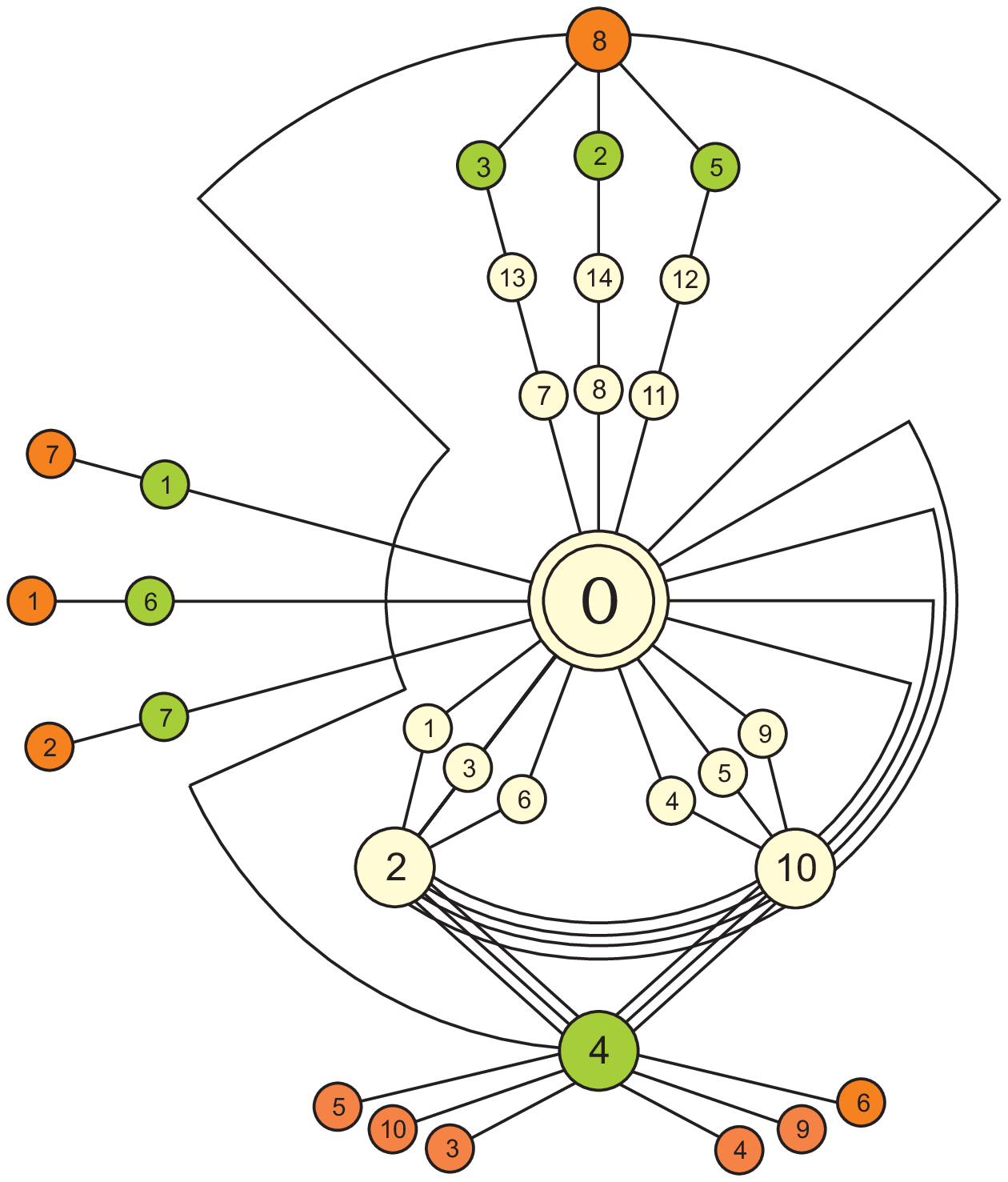}}
\vspace*{.2cm} \caption{The same as in the previous two figures for a uni-tripod (O).}
\end{figure}

Given a {\it uni}-tripod, there are 10 plane ovoids, 7 tri-tripods and 14 uni-tripods disjoint from it. Figure 10 (``frog") illustrates the configuration of 13 fans through a uni-tripod (O). We again see a remarkable pattern. One plane ovoid (8) has a special standing as there pass four different fans through it. Similarly, one tri-tripod (4) and two uni-tripods (2 and 10) have also a special footing, as each of them is shared by seven distinct fans. One sees here, however, a slightly different [3 + 3] + (3) + $\{3\}$ + 1 factorization; the ``1" stands for the distinguished fan $\{$8, 4, O, 2, 10$\}$. Moreover, whilst omitting the fixed/central ovoid in the previous two cases leaves us with only two kinds of ovoids, this case yields all the three types.

What is the nature of the three distinguished fans? This is quite easy to spot. The nine points of a {\it plane} ovoid can always be partitioned into three pairwise disjoint {\it tri}centric triads. Since a plane ovoid contains 12 tricentric triads, there exist {\it four} different ways such a partitioning can be done. Given any of them, the nine centers of the triads generate a (unique) {\it tripod}; the plane ovoid and its four associated tripods form a distinguished fan of ovoids.
In all the three above-discussed cases, the distinguished fan is {\it the only} one sharing with {\it each} of the remaining twelve at least one additional ovoid apart from the common one, and thus fully deserves its name.

\section{Conclusion}
We have furnished several examples showing that the Veldkamp space
of GQ(4,\,2) is not a (partial) linear space. This is in a sharp
contrast with the case of the dual, GQ(2,\,4), whose Veldkamp
space is a linear space, being isomorphic to PG(5,\,2)
\cite{vs24}. We surmise that this fact, among other things, might
also contain an important clue why it is GQ(2,\,4),  not
GQ(4,\,2), which is relevant for a particular black-hole/qubit
correspondence \cite{gq24}.

\section*{Acknowledgements}
This work was conceived within the framework of the Cooperation
Group ``Finite Projective Ring Geometries: An Intriguing Emerging
Link Between Quantum Information Theory, Black-Hole Physics, and
Chemistry of Coupling" at the Center for Interdisciplinary
Research (ZiF), University of Bielefeld, Germany, being completed
with the support of the VEGA grant agency, projects Nos.
2/0092/09, 2/0098/10 and 2/7012/27. I am extremely grateful to
Prof. Hans Havlicek (Vienna University of Technology) for
providing me with an easy-to-use software for generating the first seven
figures. I also thank Richard Green (University of Colorado,
Boulder) for motivating combinatorial remarks and Petr Pracna (J. Heyrovsk\' y Institute of Physical Chemistry, Prague)
for providing me with the electronic versions of the last three figures.

\end{document}